\documentclass[aps,prb,twocolumn,superscriptaddress,showpacs]{revtex4}
\usepackage{eurosym}
\usepackage{amsfonts}
\usepackage{amssymb}
\usepackage{amsmath}
\usepackage{graphicx}
\usepackage{color}
\usepackage{chapterbib}
\usepackage{threeparttable}
\usepackage{CJK}
\usepackage{sidecap}
\usepackage[hypertex,dvipdfm,colorlinks=true,urlcolor=blue,linkcolor=blue,citecolor=blue]{hyperref}
\newcommand*{\citen}{}
\DeclareRobustCommand*{\citen}[1]{%
  \begingroup
    \romannumeral-`\x 
    \setcitestyle{numbers}%
    \cite{#1}%
  \endgroup   }
\begin{document}
\title{Manipulation of type-I and type-II Dirac points in PdTe$_2$ superconductor by external pressure}
\author{R. C. Xiao}
\affiliation{Key Laboratory of Materials Physics, Institute of Solid
State Physics, Chinese Academy of Sciences, Hefei 230031, China}
\affiliation{University of Science and Technology of China, Hefei 230026, China}

\author{P. L. Gong}
\affiliation{Department of Physics, Southern University of Science and Technology, Shenzhen 518055, China}
\author{Q. S. Wu}
\affiliation{Theoretical Physics and Station Q Zurich, ETH Zurich, Zurich 8093, Switzerland}

\author{W. J. Lu}
\email{wjlu@issp.ac.cn}
\affiliation{Key Laboratory of Materials Physics, Institute of Solid State Physics, Chinese Academy of
Sciences, Hefei 230031, China}

\author{M. J. Wei}
\affiliation{Key Laboratory of Materials Physics, Institute of Solid
State Physics, Chinese Academy of Sciences, Hefei 230031, China}
\affiliation{University of Science and Technology of China, Hefei 230026, China}

\author{J. Y. Li}
\affiliation{Key Laboratory of Materials Physics, Institute of Solid
State Physics, Chinese Academy of Sciences, Hefei 230031, China}
\affiliation{University of Science and Technology of China, Hefei 230026, China}

\author{H. Y. Lv}
\affiliation{Key Laboratory of Materials Physics, Institute of Solid
State Physics, Chinese Academy of Sciences, Hefei 230031, China}
\author{X. Luo}
\affiliation{Key Laboratory of Materials Physics, Institute of Solid
State Physics, Chinese Academy of Sciences, Hefei 230031, China}
\author{P. Tong}
\affiliation{Key Laboratory of Materials Physics, Institute of Solid
State Physics, Chinese Academy of Sciences, Hefei 230031, China}
\author{X. B. Zhu}
\affiliation{Key Laboratory of Materials Physics, Institute of Solid
State Physics, Chinese Academy of Sciences, Hefei 230031, China}
\author{Y. P. Sun}
\email{ypsun@issp.ac.cn}
\affiliation{Key Laboratory of Materials Physics, Institute of Solid
State Physics, Chinese Academy of Sciences, Hefei 230031, China}
\affiliation{High Magnetic Field Laboratory, Chinese Academy of
Sciences, Hefei 230031, China}
\affiliation{Collaborative Innovation Center of Microstructures,
Nanjing University, Nanjing 210093, China }

\begin{abstract}
A pair of type-II Dirac cones in PdTe$_2$ was recently predicted by theories and confirmed in experiments, making PdTe$_2$ the first material that processes both superconductivity and type-II Dirac fermions. In this work, we study the evolution of Dirac cones in PdTe$_2$ under hydrostatic pressure by the first-principles calculations. Our results show that the pair of type-II Dirac points disappears at 6.1 GPa. Interestingly, a new pair of type-I Dirac points from the same two bands emerges at 4.7 GPa. Due to the distinctive band structures compared with those of PtSe$_2$ and PtTe$_2$, the two types of Dirac points can coexist in PdTe$_2$ under proper pressure (4.7-6.1 GPa). The emergence of type-I Dirac cones and the disappearance of type-II Dirac ones are attributed to the increase/decrease of the energy of the states at $\Gamma$ and $A$ point, which have the anti-bonding/bonding characters of interlayer Te-Te atoms. On the other hand, we find that the superconductivity of PdTe$_2$ slightly decreases with pressure. The pressure-induced different types of Dirac cones combined with superconductivity may open a promising way to investigate the complex interactions between Dirac fermions and superconducting quasi-particles.
\end{abstract}
\maketitle
\section{Introduction}
Topological (Dirac/Weyl) semimetals\cite{RN765,RN812} are new topological states of three-dimensional (3D) quantum matters, different from the topological insulators. In Dirac/Weyl semimetals, the linear band crossings are the fourfold/twofold degenerated points, whose low-energy excitations are the massless Dirac/Weyl fermions corresponding to the counterparts in the high-energy physics. By breaking the inversion symmetry or time-reversal symmetry, one Dirac fermion will transform into two Weyl ones with opposite chiralities in the Brillouin zone (BZ). In recent years, the discoveries of Dirac semimetals (such as Na$_3$Bi\cite{RN1051,RN1121} and Cd$_3$As$_2$\cite{RN1123,RN1130,RN1125}) and Weyl semimetals (such as Y$_2$Ir$_2$O$_7$,\cite{RN257} HgCr$_2$Se$_4$\cite{RN1187} and TaAs\cite{RN1132,RN1131,RN1133}) in theories and/or experiments made Dirac and Weyl fermions widely concerned.

The Dirac/Weyl semimetals that directly correspond to the counterparts in the high-energy physics are usually called type-I Dirac/Weyl semimetals. However unlike the high-energy physics, the restriction of Lorentz invariance is not necessary in the condensed matter physics. Recently a great deal of attention was paid to look for new topological quasi-particles beyond the direct counterparts in the high-energy physics.\cite{RN1327} Soluyanov \textit{et al.}\cite{RN762} proposed a new type of Weyl fermion (type-II) that contacts the bulk electron and hole pockets in the condensed matter systems. Afterwards, many type-II Weyl semimetals were discovered, such as WTe$_2$,\cite{RN762,RN1135,RN1137,RN1138} MoTe$_2$,\cite{RN206,RN991,RN1134} Mo$_x$W$_{1-x}$Te$_2$,\cite{RN992,RN656} Ta$_3$S$_2$,\cite{RN1182} TaIrTe$_4$\cite{RN1180} and LaAlGe.\cite{RN1181} The type-II Weyl semimetals show exotic properties different from the type-I ones, such as direction dependent chiral anomaly,\cite{RN762,RN1118,RN1183} enhanced superconductivity,\cite{RN1114,RN1058} anti-chiral effect of the chiral Landau level,\cite{RN1184} and novel quantum oscillations.\cite{RN1185}

Researchers are also trying to find the type-II Dirac fermions in the condensed matters. Recently, Huang \textit{et al.}\cite{RN980} predicted that the type-II Dirac fermions protected by $C_3$ rotational symmetry can exist in the PtSe$_2$ family materials, and similar proposals were also put forward by Le \textit{et al.}\cite{RN1179} and Chang \textit{et al.}\cite{RN1178} in the KMgBi and VAl$_3$ family materials, respectively. Following Huang's predictions, the evidences of type-II Dirac cones in PtTe$_2$,\cite{RN1011,RN1107} PdTe$_2$\cite{RN1107,RN1028,RN1111} and PtSe$_2$\cite{RN1107,RN1142} were soon characterized in angle-resolved photoemission spectroscopy (ARPES) experiments by different groups. Similar to the type-II Weyl cones, the type-II Dirac cones are strongly tilted in some specific directions. Novel physical properties different from those in the standard type-I Dirac semimetals are expected in the type-II Dirac semimetals.\cite{RN980,RN1178} Interestingly, PdTe$_2$ is also a superconductor with  transition temperature ($T_C$) about 1.7-2.0 K.\cite{RN1028,RN750,RN1047} The coexistence of superconductivity and type-II Dirac points in PdTe$_2$ makes it significantly different from other members of PtSe$_2$ family materials, which could provide a possible platform to explore the interplay between superconducting quasi-particles and Dirac fermions.\cite{RN1107,RN1028,RN1111}

Pressure can drive the topological phase transitions in the topological materials and can also assist to comprehend the nature of topological states at ambient pressure. In this work, we focus on studying the evolution of the Dirac points and superconductivity in PdTe$_2$ under hydrostatic pressure by the first-principles calculations. Our results show that the pair of type-II Dirac points disappears at 6.1 GPa, while a new pair of type-I Dirac points emerges at 4.7 GPa. Under the pressure of 4.7-6.1 GPa, the type-II and type-I Dirac cones coexist. The evolution of the two types of Dirac cones can be understood by the bonding and anti-bonding characters near the Dirac cones. We also find that the superconductivity slowly decreases with the increase of pressure, meaning that there are abundant topological transitions together with superconductivity in PdTe$_2$ under the external pressure.

\begin{figure}
\begin{flushleft}
\centering
\includegraphics[width=1\columnwidth]{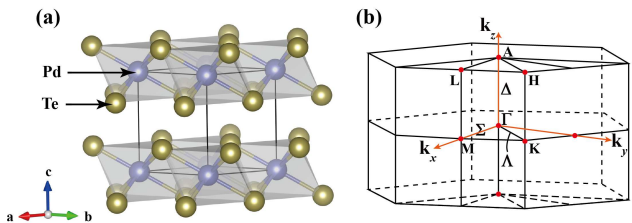}
\caption{(a) Crystal structure and (b) Brillinon zone of PdTe$_2$.}
\label{Cry_structrue}
\end{flushleft}
\end{figure}

\section{Method}
The first-principles calculations based on density functional theory (DFT) were performed using QUANTUM-ESPRESSO package.\cite{RN82} The ultrasoft pseudo-potentials and the general gradient approximation (GGA) according to the PBE functional were used. The energy cutoff of the plane wave (charge density) basis was set to 50 Ry (500 Ry). The BZ was sampled with a $12\times12\times8$ mesh of $k$-points. The Methfessel-Paxton Fermi smearing method with a smearing parameter of $\sigma = 0.02$ Ry was used. The lattice constants and ions were optimized using Broyden-Fletcher-Goldfarb-Shanno (BFGS) quasi-newton algorithm. All of the band structure calculations are cross-checked by the VASP codes,\cite{RN1434,RN1433} and the results are consistent with each other.
\section{ Results and Discussion }

\begin{figure}
\begin{flushleft}
\centering
\includegraphics[width=1\columnwidth]{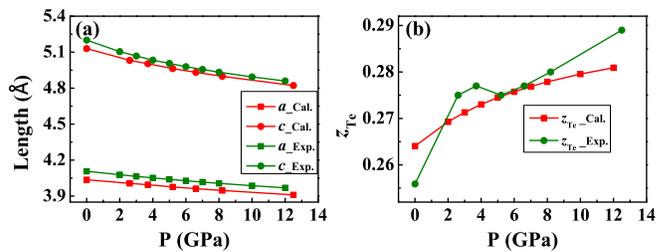}
\caption{(a) Lattice constants and (b) $z_{Te}$ of PdTe$_2$ under pressure. Experimental data are from Ref.~\citen{RN1018}.}
\label{pressure-lattice}
\end{flushleft}
\end{figure}

PdTe$_2$ belongs to transition metal dichalcogenides (TMDCs), and has a layered structure with space group P\=3m1 ($1T$-CdI$_2$ type structure) with one Pd atom and two Te atoms located at (0, 0, 0) and (1/3, 2/3, $\pm z_{Te}$) sites, respectively (Fig. \ref{Cry_structrue}(a)). One Pd atom and the nearest six Te atoms compose an octahedron. The lattice constant ratio $c/a$ ($\approx$1.27) is the smallest among the iso-structural TMDCs,\cite{RN665} leading to the octahedra largely distorted. The special $k$ points and the paths of BZ are shown in Fig. \ref{Cry_structrue}(b), and the corresponding symmetries are summarized in Table S1 in the Supplemental Material.\cite{SM_URL} The calculated lattice constants $a$ and $c$ at ambient pressure are slightly overestimated by about $1.76 \%$ and $1.36 \%$ respectively, due to the underestimation of bond strength in GGA. No structural phase transition was found up to 27 GPa in experiment.\cite{RN1018} The optimized lattice constants under pressure are shown in Fig. \ref{pressure-lattice}.

\begin{figure}
\begin{flushleft}
\centering
\includegraphics[width=0.9\columnwidth]{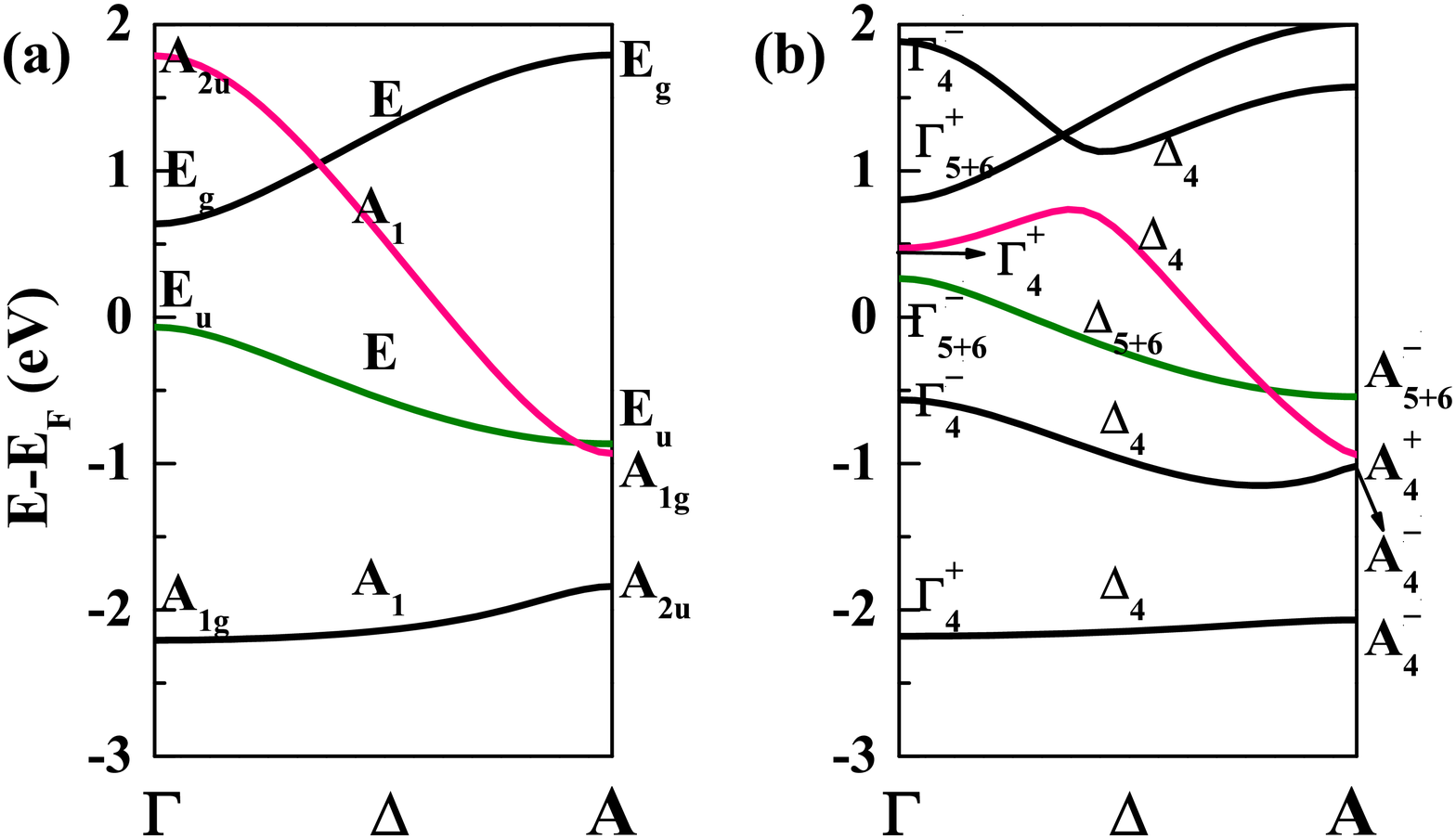}
\caption{Band structures of PdTe$_2$ (a) without and (b) with SOC along $\Gamma$-$A$ direction at ambient pressure. The irreducible representations are indicated. The band symmetries are also denoted near each bands, and the IRs are introduced in Table S2-S5 in the Supplemental Material.\cite{SM_URL}}
\label{band-0GPa}
\end{flushleft}
\end{figure}

\begin{figure*}
\centering
\includegraphics[width=0.85\textwidth]{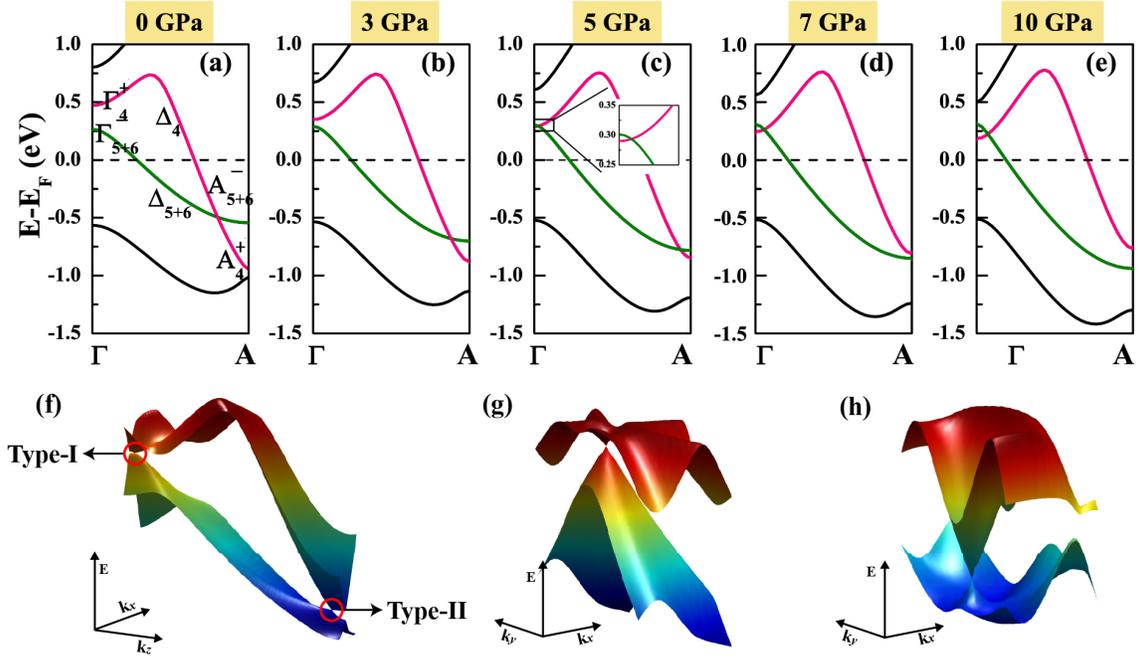}
\caption{(a-e) Evolution of the Dirac points of PdTe$_2$ from 0 GPa to 10 GPa. 3D band structures at 5 GPa on the (f) $k_x$-$k_z$ plane, (g) $k_z= 0.012$ $2\pi/c$ plane around the type-I Dirac cone and (h) $k_z$= 0.468 $2\pi/c$ plane around the type-II Dirac cone. The band structures at 5 GPa along the in-plane and out-plane directions are shown in Fig. S3 in the Supplemental Material.\cite{SM_URL} }
\label{band_evalution}
\end{figure*}

\begin{figure}
\centering
\includegraphics[width=0.98\columnwidth]{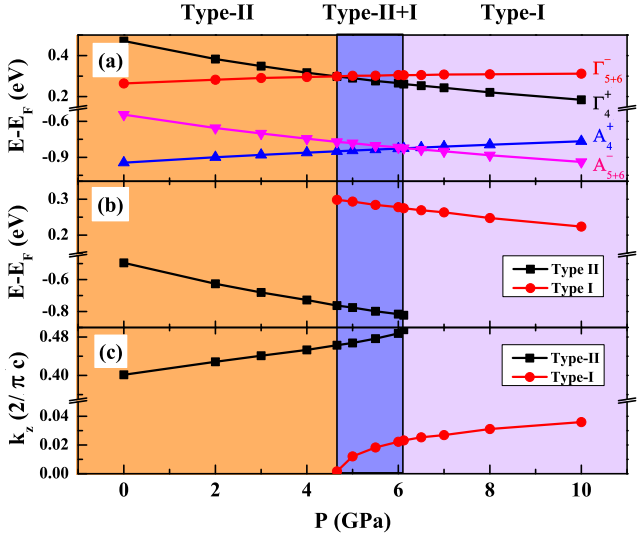}
\caption{(a) Energy evolution of the $\Gamma_4^+$, $\Gamma_{5+6} ^-$, $A_4^+$ and $A_{5+6} ^-$ states of PdTe$_2$ under the pressure. (b) Energy of and (c) positive positions of the type-II and type-I Dirac points under pressure. The regions decorated by colors stand for the existing ranges of Dirac cones.}
\label{state_energy}
\end{figure}

The band structures along $\Gamma$-$A$ direction at ambient pressure without/with spin orbit coupling (SOC) effects are shown in Fig. \ref{band-0GPa}. Without SOC, the $E$ band (denoted by green color) and $A_1$ (denoted by red color) band, which belong to the two different irreducible representations (IRs), cross each other near the $A$ point. The $E$ band is mainly composed by the Te-$p_x$+$p_y$ orbitals, while the $A_1$ band is mainly composed by the Te-$p_z$ orbitals (see Fig. S1 in the Supplemental Material\cite{SM_URL}), leading to the $A_1$ band (red) more dispersive than the $E$ band (green). When the SOC is included, the band structures change dramatically as shown in Fig. \ref{band-0GPa}(b). All the $A_1$ bands transform into the two dimensional (2D) IR $\Delta_4$ bands, and all the $E$ bands split into $\Delta_4$ (2D IR) and $\Delta_{5+6}$ bands (the $\Delta_5$ (1D IR) and $\Delta_6$ (1D IR) bands are degenerated along the $\Gamma$-$A$ direction). The band crossing near the $A$ point is also inevitable with SOC, since the $\Delta_4$ (red) and $\Delta_{5+6}$ (green) bands belong to different 2D IRs. The strongly dispersive band $\Delta_4$ crosses the less dispersive band $\Delta_{5+6}$, resulting in a titled Dirac cone along $k_z$, which is the type-II Dirac cone previously proposed by theories and verified in ARPES experiments.\cite{RN1107,RN1028,RN1111} The type-II Dirac cone protected by $C_3$ rotational symmetry is tilted along $\Gamma$-$A$ direction but untilted on the $k_x$-$k_y$ plane. According to the crystal symmetry, there is another type-II Dirac point located at the opposite position of the BZ, thus there is a pair of type-II Dirac cones. The location of the pair of type-II Dirac points in PdTe$_2$ is closer to the Fermi energy ($E_F$) than those of PtSe$_2$ and PdTe$_2$.\cite{RN980}

Because the $\Delta_4$ band (red) and the upper $\Delta_4$ band (black) belong to the same IR, as a result a band inversion at $\Gamma$ point instead of a band crossing at $\Gamma$-$A$ direction occurs (Fig. \ref{band-0GPa}(b)). Topological surface state located at $\overline{\Gamma }$ deeply below $E_F$ at -1.75 eV was observed in experiments, which was considered to be due to the band inversion between the two lowest two $\Delta_4$ bands in Fig. \ref{band-0GPa}(b).\cite{RN1107,RN1111,RN750}

The $A_4^+$ state moves upward and while the $A_{5+6}^-$ state moves downward relative to $E_F$ with pressure (Figs. \ref{band_evalution}(a)-(e)), thus this pair of type-II Dirac cones gradually disappears near the $A$ point. The $\Gamma_4^+$ state moves downward, while the $\Gamma_{5+6}^-$ state moves upward with pressure, so a new pair untitled Dirac points \emph{i.e.} type-I Dirac points, from the same two bands emerges near the $\Gamma$ point. The energy evolution of the $\Gamma_4^+$, $\Gamma_{5+6}^-$, $A_4^+$ and $A_{5+6}^-$ states under pressure is summarized in Fig. \ref{state_energy}(a), from which we can see that the pair of type-I Dirac cones emerges at 4.7 GPa, and the pair of type-II Dirac cones disappears at 6.1 GPa. Under the pressure of 4.7-6.1 GPa, the two kinds of Dirac points coexist (for instance the 3D band structures at 5 GPa in Figs. \ref{band_evalution}(f)-(h)). When pressure further increases, it remains only the pair of type-I Dirac cones. The energy and positions of the two kinds of Dirac points are shown in Figs. \ref{state_energy}(b) and (c). The type-II and type-I Dirac points locate below and above the $E_F$ respectively, and both move downward with pressure. Under higher pressure (about 30 GPa) the pair of type-I Dirac points will move to $E_F$. In the coexistence range, the type-II and type-I Dirac points are well separated in the BZ, which can be easily detected in experiments.

\begin{figure}
\centering
\includegraphics[width=1\columnwidth]{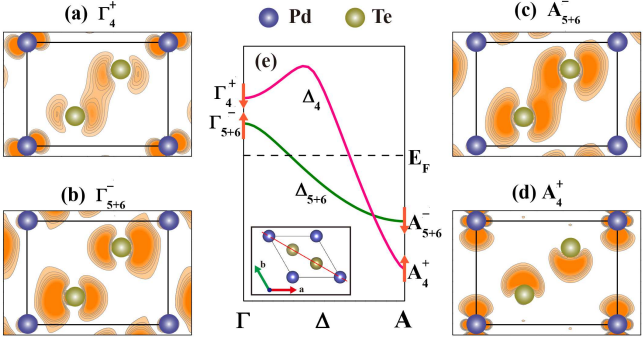}
\caption{Charge density of the (a) $\Gamma_4^+$, (b) $\Gamma_{5+6}$, (c) $A_{5+6}^-$ and (d) $A_4^+$ states of PdTe$_2$ in the (110) plane at ambient pressure. The arrows in (e) denote the moving directions under pressure, and the inset illustrates the (110) plane.}
\label{Charge_density}
\end{figure}

To reveal why the type-II Dirac points disappear and type-I Dirac points emerge, the charge density of the $\Gamma_4^+$, $\Gamma_{5+6}^-$, $A_4^+$ and $A_{5+6}^-$ states in real space was calculated and shown in Fig. \ref{Charge_density}. The interlayer Te-Te atoms show a bonding character in the $\Gamma_4^+$ (Fig. \ref{Charge_density}(a)) and $A_{5+6}^-$ (Fig. \ref{Charge_density}(c)) states, while the interlayer Te-Te atoms show an anti-bonding character in the $\Gamma_{5+6}^-$ (Fig. \ref{Charge_density}(b)) and $A_4^+$ (Fig. \ref{Charge_density}(d)) states. 
Since the interlayers bind to each other via a weak van der Waals interaction, and the Pd atom orbitals are localized and the Te atom orbitals are extended in the $\Gamma_4^+$ and $A_4^+$ states, the Te atom orbitals are easier to be affected by pressure. Pressure decreases the atom distance and enhances the atom interaction, leading to the energy of the bonding states moving downward, and the energy of the anti-bonding states moving upward relative to $E_F$. The opposite bonding character of $\Gamma_4^+$ and $\Gamma_{5+6}^-$ states makes their energy move in opposite directions with pressure, and so do the $A_4^+$ and $A_{5+6}^-$ states. The similar bonding and anti-bonding character of states make the topological phase transitions or metal-insulator transitions in Bi$_2$Se$_3$ \cite{RN1150}and phosphorene\cite{RN1259,RN1153} under pressure. Just as the statement in Ref.~\citen{RN1107}, the formation of the type-II Dirac cones is very general: just due to the different dispersion of two IRs bands. Pressure can manipulate the band inversion and band dispersion, so more type-II Dirac and type-I Dirac materials are expected to be found in TMDCs under pressure.

\begin{figure}
\centering
\includegraphics[width=1\columnwidth]{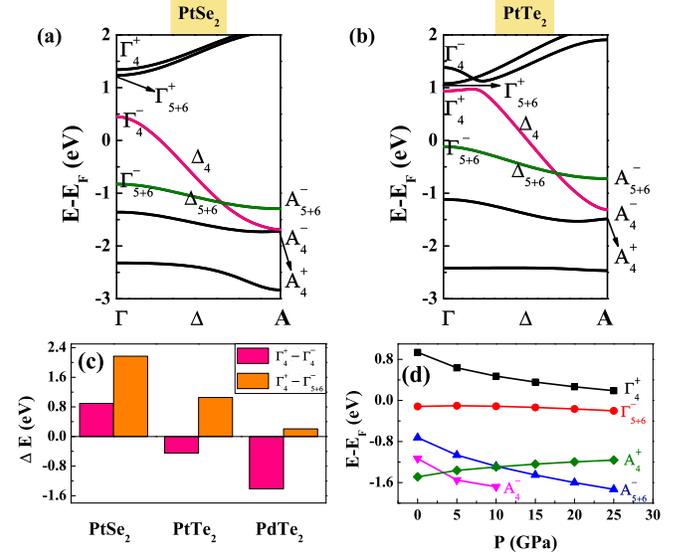}
\caption{Band structures of (a) PtSe$_2$ and (b) PtTe$_2$ along $\Gamma$-$A$ direction. (c) Inversion energy of $\Gamma_4^+$ and $\Gamma_4^-$ states and energy difference of $\Gamma_4^+$ and $\Gamma_{5+6}^-$ states for three materials at ambient pressure. (d) Energy evolution of $\Gamma_4^+$, $\Gamma_{5+6}^-$, $A_4^+$ and $A_{5+6}^-$ states of PtTe$_2$ under pressure. The $A_4^+$ and $A_4^-$ states inverse with pressure (see Fig. S5).}
\label{PtSe2-PtTe2}
\end{figure}

Due to the same crystal structure, PtSe$_2$ and PdTe$_2$ show similar band structures to PdTe$_2$ (see Figs. \ref{PtSe2-PtTe2}(a) and 7(b)). However the differences also exist. The $\Gamma_4^+$ and $\Gamma_4^-$ states inverse in PtTe$_2$ (Fig. \ref{PtSe2-PtTe2}(b)) and PdTe$_2$ (Fig. \ref{band-0GPa}(b)), but they do not inverse in PtSe$_2$ (Fig. \ref{PtSe2-PtTe2}(a)). The inversion energy of the $\Gamma_4^+$ and $\Gamma_4^-$ states (1.41 eV) is the largest, and the energy difference of $\Gamma_4^+$ and $\Gamma_{5+6}^-$ states (0.21 eV) is the smallest in PdTe$_2$ (see Fig. \ref{PtSe2-PtTe2}(c)). Due to the distinctive band structures, two kinds of Dirac points are easier to coexist in PdTe$_2$ compared with the cases of PtSe$_2$ and PtTe$_2$. Taking PtTe$_2$ as an example, the evolution of band structures under external pressure in PtTe$_2$ (see Fig. S5 in the Supplemental Material\cite{SM_URL}) is similar to those of PdTe$_2$. The $ A_4^+$/$A_{5+6}^-$ state of PtTe$_2$ moves upward/downward with pressure, and the pair of type-II Dirac points gradually disappears at about 10 GPa (see Fig. \ref{PtSe2-PtTe2}(d)). Meanwhile the $\Gamma_4^+$/$\Gamma_{5+6}^-$ state moves downward/upward with pressure, and a new pair of type-I Dirac points will emerge at higher pressure. But the type-II and type-I Dirac points do not coexist in PtTe$_2$ under pressure. Similar evaluation of Dirac points in PtSe$_2$ is summarized in the Supplemental Material.\cite{SM_URL}

\begin{figure}
\centering
\includegraphics[width=0.8\columnwidth]{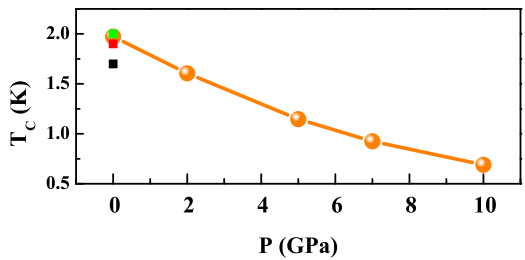}
\caption{Calculated $T_C$ of PdTe$_2$ under pressure (orange balls). The squares at ambient pressure indicate the experimental detected values. The black square is from Refs.~\citen{RN750} and ~\citen{RN748}, the red one is from Ref.~\citen{RN1028} and the green one is from Ref.~\citen{RN1047}.}
\label{TC_P}
\end{figure}

We also evaluated the phonon and electron-phonon coupling of PdTe$_2$ under pressure. The phonon branches become more dispersive and shift to higher frequency with pressure. The calcultaed $T_C$ decreases slowly with the nearly linear rate of 0.13 K/GPa from 1.97 K at ambient pressure to 0.69 K at 10 GPa (see Fig. \ref{TC_P}). The decrease of electronic density of states at $E_F$ ($N(E_F)$) and the blueshift of phonon density of states ($F(\omega)$) contribute to the decrease of $T_C$. Though $T_C$ decreases monotonously with pressure, it is still in an experimentally detectable range. Especially, the $T_C$ is above 1.0 K in the coexistence range of two types of Dirac cones. The detailed discussions are in the Supplemental Material.\cite{SM_URL}

The intrinsic topology property can drive the conventional superconductivity into nontrivial $p$-wave-like unconventional topological superconductivity via conventional phonon mediated mechanism,\cite{RN1379,RN1422} like doped Bi$_2$Se$_3$\cite{RN1379,RN1415,RN1312,RN1318,RN1380} and pressed Cd$_3$As$_2$.\cite{RN1308,RN1314} Thus the issue whether PdTe$_2$ is a topological superconductor needs to be further investigated. The electron and hole pockets near the type-II Dirac points provide plentiful density of states than those of the type-I Dirac points due to the titled cones. Like the type-II Weyl points,\cite{RN1114,RN1058} the type-II Dirac points are favorable for both superconducting and topological superconducting carrier ratios.\cite{RN1028}  As for the coexistence of type-I and type-II Dirac points in PdTe$_2$, more superconducting carrier ratios could be expected, and more interesting superconductivity properties should appear, which is deserved to be investigated further.

In addition, because the lattice constants under pressure optimized by GGA show about 2$\%$ systematic overestimation with the experimental data,\cite{RN1018} we calculated the band structures using the experimental data to estimate the overestimation (see Fig. S11 in the Supplemental Material\cite{SM_URL}). The pairs of type-II and type-I Dirac cones can still coexist under proper pressure when the experimental lattice constants are adopt, and the transition pressure of the Dirac points may be smaller than that using the optimized structure data.

\section{Conclusion}
The type-I and type-II Dirac points in PdTe$_2$ can be tuned by applying the external pressure. The pair of type-II Dirac points disappears with pressure at 6.1 GPa, while a new pair of type-I Dirac points stemming from the same bands emerges at 4.7 GPa. The two types of Dirac points can coexist under proper pressure (4.7-6.1 GPa). The increase of the $A_4^+$ state energy and the decrease of the $A_{5+6}^-$ state energy make the pair of type-II Dirac points gradually disappear. The increase of the $\Gamma_{5+6}^-$ state energy and the decrease of the $\Gamma_4^+$ state energy make the type-I Dirac points emerge. The decrease/increase of the ($\Gamma_4^+$ and $A_{5+6}^-$)/($\Gamma_{5+6}^-$ and $A_4^+$) states energy is attributed to the bonding/anti-bonding character of interlayer Te-Te atoms. Due to the distinctive band structures of PdTe$_2$, the type-II and type-I Dirac points from two same bands can coexist in PdTe$_2$ under appropriate pressures. The superconductivity weakens monotonously with pressure with the average rate of 0.13 K/GPa due to the decrease of $N(E_F)$ and the blueshift of $F(\omega)$, but still in an experimentally detectable range. Given the abundant topological transitions and superconductivity under pressure, PdTe$_2$ under pressure may be an interesting topological Dirac material. Further experimental verification and theory studies need to be addressed.

\begin{acknowledgements}
This work was supported by the National Key Research and Development Program of China under Contract No. 2016YFA0300404, the National Nature Science Foundation of China under Contract Nos. 11674326, 11404340, 11404024 and U1232139, Key Research Program of Frontier Sciences of CAS (QYZDB-SSW-SLH015) and Hefei Science Center of CAS (2016HSC-IU011). The calculations were partially performed at the Center for Computational Science, CASHIPS. Quansheng Wu was supported by Microsoft Research, and the Swiss National Science Foundation through the National Competence Centers in Research MARVEL and QSIT. We also acknowledge Dr. Chi-Ho Cheung in National Taiwan University for the useful discussions.
\end{acknowledgements}

\end {document}